\documentclass[pre,twocolumn,superscriptaddress,floatfix,amsmath,a4paper]{revtex4-2}
\usepackage{graphicx}
\usepackage[english]{babel}
\usepackage{amsmath,amssymb}
\usepackage[utf8]{inputenc}
\usepackage{psfrag}
\usepackage{bm}
\usepackage{xcolor}
\usepackage{hyperref}
\usepackage{appendix}
\usepackage[normalem]{ulem}

\begin{document}

\title{Moderate immigration may promote a peak of cooperation among natives }
\author{Alessandra F. Lütz}
\email{sandiflutz@gmail.com}
\affiliation{Departamento de Física, Universidade de Minas Gerais, 31270-901, Belo Horizonte MG, Brazil}
\author{Marco A. Amaral}
\affiliation{Instituto de Humanidades, Artes e Ciências, Universidade Federal do Sul da Bahia, 45988-058, Teixeira de Freitas BA, Brazil}
\author{Lucas Wardil}
\affiliation{Departamento de Física, Universidade de Minas Gerais, 31270-901, Belo Horizonte MG, Brazil}
\date{\today}

\begin{abstract}
In a world of hardening borders, nations may deprive themselves of enjoying the benefits of cooperative immigrants.  Here, we analyze the effect of efficient cooperative immigrants on a population playing public goods games. We considered a population structured on a square lattice with individuals playing public goods games with their neighbors. The demographics are determined by stochastic birth, death, and migration. The strategies spread through imitation dynamics. Our model shows that cooperation among natives can emerge due to social contagion of good role-model agents that can provide better quality public goods. Only a small fraction of efficient cooperators, among immigrants, is enough to trigger cooperation across the native population. We see that native cooperation achieves its peak at moderate values of immigration rate. Such efficient immigrant cooperators act as nucleation centers for the growth of cooperative clusters, that eventually dominate defection.
\end{abstract}


\maketitle

\section{Introduction}
International migration plays a major role in our globalized world~\cite{migration_climate}. Highly skilled individuals are often needed and most welcomed in countries with a lack of specialized labor force~\cite{immigration-selection}. Also, the international refugee crisis is becoming more common each year, motivated by civil wars, climate changes, disease, and many other factors~\cite{AbouChakra2018, Vicens2018, Curry2020, Couto2020}. There is no doubt that the economic growth in many countries has received a valuable contribution from both natives (i.e. native-born citizens) and immigrants~\cite{immigration-economicgrowth}.

Much of the greatest challenges humanity has been facing recently are rooted in the coordination of individual actions toward the common good~\cite{Pennisi2005, Capraro2018}. Thus, different areas have explored this matter, and there is extensive literature focused on the emergence of cooperation (e.g.~\cite{rand_tcs13, Perc2017, buchan_pnas09, Gomez-Gardenes2007, Nowak2011a, Hauert2005, CaMa18, Hofbauer1998, galam_ijmpc08}). 
Because the environmental pressure is creating urgent demands for more altruism, a highly welcomed kind of immigrant turns out to be the cooperative one. Super-cooperative immigrants could even create a positive synergistic environment, promoting cooperation among natives. Some famous examples are humanitarian organizations such as the Red Cross, the World Food Program, and the Doctors Without Borders among many others~\cite{Beigbeder91,banatvala2000public}. 
Those organizations send human resources to aid developing communities as well as places hit by disasters or civil wars.
Such initiatives can strongly act as seeds for widespread cooperation, even when the so said super-cooperators are not permanent immigrants. For immigrants who move permanently into a community, several empirical studies suggest that their presence may help boost innovation and benefit such community~\cite{HuGa10, kerr2013us, ozgen2012immigration}.

Human altruism is rooted in both genetic and cultural bases~\cite{boyd:culture}. While selfishness generally permeates all societies at some level, moral rules have co-evolved with cooperative behaviors, yielding highly stable cooperative societies~\cite{Nowak2011a, Perc2017}. Individuals that are raised in altruistic societies often embrace a highly cooperative phenotype and hardly deviate from it, suffering from guilty and other negative emotions if deviation happens~\cite{jacquet}. When individuals nurtured in altruistic societies migrate to other places, their traits will likely remain the same, which allows them to influence natives' behaviors. Their offspring, however, may embrace the new society's culture and not behave like their parents~\cite{unaosili}. To investigate such social dynamics, here we use evolutionary game theory to analyze the arrival of both egotistic and role-model altruistic immigrants into a community. 

In the context of evolutionary game theory, cooperation is often modeled as a public goods game~\cite{Perc2017, szolnoki_epl10}. Typically, $n$ individuals are asked to invest a fixed resource amount of $c$ to a common pool. Each contribution is multiplied by a factor $r$, summed up, and the total is divided equally among all $n$ individuals, independently of their chosen strategy. The multiplication factor quantifies how efficiently individual resources aggregate to create public goods. This may depend on the resource quality of each contribution or the agent's capacity to increase the value of the common pool. Because defectors -- individuals that do no contribute --  earn more than cooperators, they become wealthier, and cooperation does not thrive. However, if mechanisms like punishment, reward, or spatial structure are present, cooperation may have a chance to flourish~\cite{Krawczyk2018, Javarone2016d, Wardil2017, Fang2019, Perc2016, Szabo2007, perc_bs10, perc_jrsi13, Vainstein2001a, vainstein_jtb07, Wang2012,vainstein_pa14, Zhao2020}. 

To investigate the role that super-cooperative immigrants may have on the native-born cooperation dynamics, we consider a public goods game with three types of strategies: defectors, that do not contribute; standard cooperators, that contribute $c$ and have a multiplicative factor $r$ ($1<r<n$) and efficient cooperators, that also contribute $c$, but have a multiplicative factor $\alpha r$ ($\alpha >1$). The population is structured on a square lattice and the demographics are determined by stochastic birth, death, and migration. Only the immigrants can adopt the efficient cooperation strategy. The strategies spread through imitation dynamics, such that the strategies yielding higher payoffs spread at higher rates~\cite{Szabo2007}. Here, we show that cooperation among natives reaches a peak at moderate migration rates and that even a small fraction of cooperation among the immigrants is enough to boost native cooperation. We also note that while the group size variation can promote cooperation by itself when the group is smaller than the multiplicative factor~\cite{Killingback2006}, here the efficient cooperators have a multiplicative factor greater than the group size, turning, at least for the efficient cooperators, the investment option into the most profitable decision. However, when the payoff of the efficient cooperators is compared to that of the defectors in a single game, defectors always win. The effect of demographic parameters is discussed.  Our results suggest that closing borders, to avoid exploitation by non-natives, may be counter-productive, because the positive spillover can be greater than the harms of the potential exploitation. In the next sections, we describe the model, analyze the results and provide an overall discussion of our work.

\section{The model}
We define \textit{natives} as all individuals that were born in the population. The \textit{immigrants} are those that were born outside and come to live permanently in the population. Efficient cooperation is restricted to immigrants. The assumption is that the efficiency is nurtured only in the original immigrant culture and cannot be transmitted far from the original influences. Therefore, the strategy space available for the natives is cooperation ($c$) and defection ($d$), whereas for the immigrants it is cooperation, defection, and efficient cooperation ($e$).

In a public goods game played in a group of size $n$, if $n_c$ and $n_{e}$ are, respectively, the number of standard and efficient cooperators in a given group, the payoffs of a cooperator and a defector are given by
\begin{eqnarray}
\Pi_c=\Pi_{e}&=&-1+\frac{rn_c+\alpha r n_e}{n}, \label{eq.payoffC}\\
\Pi_d&=&\frac{rn_c+\alpha r n_e}{n}, \label{eq.payoffD}
\end{eqnarray}
where we set $c=1$ for simplicity. The parameter $\alpha$ is the super-cooperator's efficiency. Note that the payoff of cooperators and efficient cooperators are identical. Also, we are assuming that efficient cooperators contribute the same amount $c$, but their ability allows a greater multiplicative factor. We stress that Eqs.~(\ref{eq.payoffC}) and~(\ref{eq.payoffD}) account only for the payoff in a single game. To obtain the total payoff, one must consider all the games a player participates in, each one with a different $n_c$ and $n_e$.

The population is structured on a square lattice of linear size $L=100$. Each site is the center of a public goods game. Thus, each individual can be part of five games: the one centered on the individual himself and the other four centered around his four nearest-neighbors. The individual payoff is the total accumulated in all games. 

The population density changes due to birth, death, and migration, while the strategy of the individuals changes due to imitation dynamics. We investigate the system using Monte Carlo simulations. If there are $n_a$ occupied sites at time $t$, then a single Monte Carlo step (MCS) in time $t$ comprises $n_a$ repetitions of one imitation, one birth, one death, and one immigration attempt.

Let us first define the imitation process. A random player $i$ is selected to imitate a random player $j$, in his neighborhood, with a probability
\begin{equation}
p_{i \rightarrow j}=max \left\{  \frac{\Pi_{j}-\Pi_{i}}{\Delta \Pi_{max}},0 \right\} 
\end{equation}
where $\Delta \Pi_{max}$ is the maximum payoff difference considering all possible combinations of allowed payoffs, which is included to normalize the probabilities. Here, $\Pi_i$ is the total payoff of player $i$, obtained by summing the payoff obtained in all games, Eqs.~(\ref{eq.payoffC}) or~(\ref{eq.payoffD}), that player $i$ participates.
Notice that the strategy space available for the natives is $\{c,d\}$, whereas for the immigrants is $\{c,d,e\}$. To allow some sort of influence of the local culture on the immigrants, we allowed the immigrants to imitate the native regular cooperation or defection, but not the other way round. For example, if a native tries to imitate the efficient cooperator strategy, he can only become a regular cooperator. Nevertheless, an efficient cooperator immigrant can imitate the cooperative strategy of a native. Notice that the regular and the efficient cooperators can be both seen as individuals whose actions are classified as cooperation.

Next, we describe the demographic part of the model. Another individual and one of its nearest neighbor sites are randomly selected. If that neighbor site is empty, the selected individual reproduces with probability $\beta$. The offspring inherits the parent's strategy, except efficient cooperators give birth to standard cooperators due to the cultural integration that the offspring is subjected to.
For the death event, another individual is randomly selected to die with probability $d=\gamma \beta$. 
Notice that, because the players can have empty sites around them, the size of public goods groups can be smaller than five.
In the immigration step, we randomly select a site and, if the site is empty, it receives an immigrant with probability $min \{1,\mu/\rho_a\}$,  where $\rho_a$ is the fraction of occupied sites and $\mu$ is the immigration coefficient. As long as $\mu \le \rho_a$, the factor $\rho_a$ guarantees that at each MCS the number of incoming immigrants is, on average, equal to $(\mu/\rho_a)(1-\rho_a)N_a=\mu(1-\rho_a)N=\mu N_v$, where $N_v$ is the number of vacant sites. Notice that is so because at each MCS both the demographic and the imitation processes are repeated $N_a$ times. Thus, the total number of incoming immigrants depends not only on the immigration coefficient $\mu$ but also on the global density. The constant influx of immigrants is comprised of a fraction $p_e$ of efficient cooperators and $1-p_e$ of defectors.

The simulations are run in a square lattice $100\times100$ starting fully occupied with natives. We consider a transient around $10^4$ to $10^5$ MCS, after which we average the measures over the last $1000$ steps. The results are further averaged over $100$ independent samples.
In the following analysis, we  distinguish five types of individuals: native cooperators ($c_0$), native defectors ($d_0$), immigrant cooperators ($c_i$), immigrant defectors ($d_i$), and immigrant efficient cooperators ($e$).

\section{Results}

\begin{figure*}[ht]

\hspace{1.5cm}$a)$\hspace{6.5cm} $b)$ \hspace{4.5cm} $c)$

\includegraphics[width=7cm]{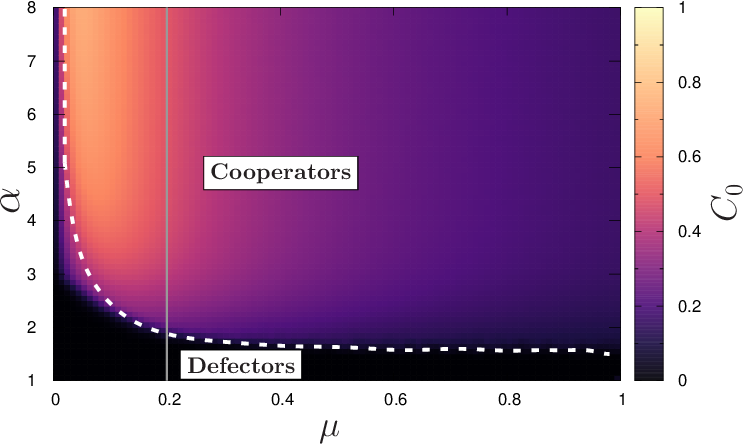}
\includegraphics[width=5.5cm]{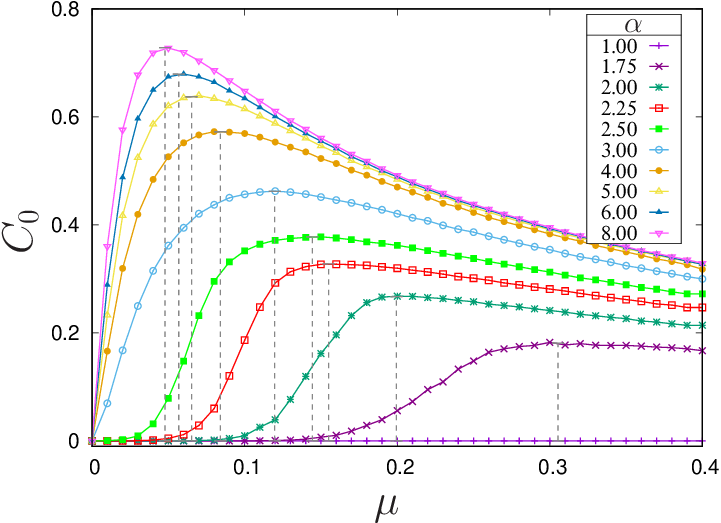}
\includegraphics[width=4.07cm]{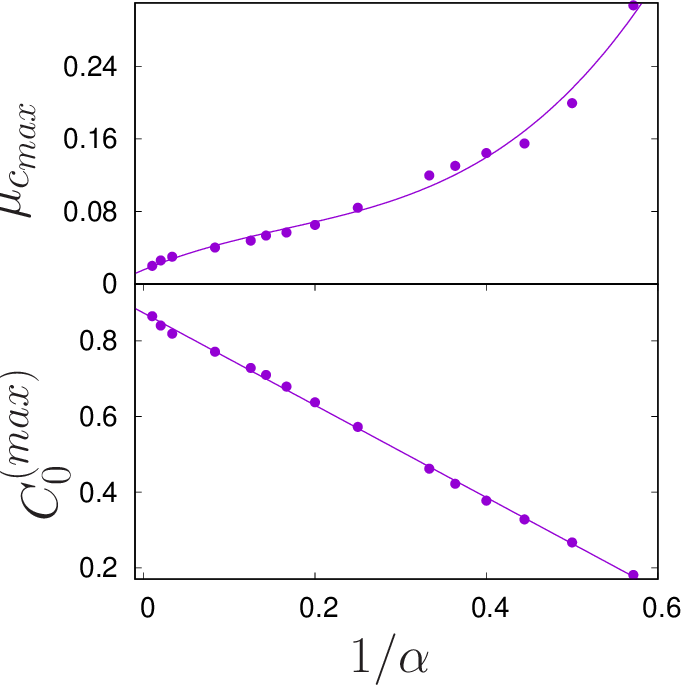}

\caption{Analysis of the prevalence of native cooperation. In ($a$) we show a heat map for the average equilibrium density of native cooperators as a function of the super-cooperator's efficiency, $\alpha$ and the immigration coefficient, $\mu$. The normalization considered for the native cooperator density is the system size. The dashed white line separates the region above which the sum of all cooperators is greater than the defectors. The grey continuous line separates the region where natives, at the left, are the majority. Notice that the grey line corresponds to $\mu\simeq\beta$, where the reproduction and immigration coefficients are approximately equal. The optimal combinations of $\alpha$ and $\mu$ for native cooperators are located in the red region, that is, small values of the immigration coefficient, $\mu$, and high levels of efficiency, $\alpha$. For most parameters considered, native cooperators are able to persist even though $r=3$. This is true even for a large immigration influx, when $\mu\to1$, although in this case, immigrants are the majority. In ($b$), we show the equilibrium density of native cooperators, $C_0$,  as a function of $\mu$ for different $\alpha$ values. Notice that each curve has a peak where $C_0$ reaches a maximum. The coordinates of those peaks $(\mu_{cma},C_0^{(max)})$ are presented as functions of $1/\alpha$ in ($c$). In all 3 figures we use $\gamma=0.005$, $\beta=0.2$ and $p_e=0.5$. Notice these results remain the same for any initial condition where the system is not empty.}
\label{fig.diagrama}
\end{figure*}

In the absence of immigration, $\mu=0$, the main parameter that allows the survival of cooperators is the multiplicative factor $r$. As $r$ increases, cooperative clusters become more and more stable in a sea of defectors, until they can survive. For our model, simulations show that cooperation is extincted for $r<3.4$.  When $r$ is that low, the benefits from clustering are not enough to face the defectors. However, if efficient cooperative immigrants are allowed into the society, i.e. $\mu>0$, even if they come along with defecting immigrants, the native cooperation can be sustained. This can be seen in Fig.~\ref{fig.diagrama}$a$, where we present the equilibrium density of native cooperators, $C_0$, as a function of the super-cooperator's efficiency, $\alpha$, and the immigration coefficient, $\mu$, in a heat map. We set $r=3$, a value where cooperation would be extinct without immigration, so as to understand the effects of efficient cooperators in the dynamics. Notice that $C_0$ is the number of native cooperators in the system normalized by the system size.
Surely, the solution where the native population is, for the most part, replaced by efficient cooperative immigrants is trivial (high $\mu$ and $\alpha$). Instead, ideally one wants to promote local cooperation through a minimal amount of incentives. Hence, allowing just a small fraction of immigrants from highly cooperative societies may create a sustainable environment even for native cooperation. This scenario corresponds to the bright red region of the heat map, where we see a small fraction of super-cooperative immigrants being able to greatly boost the native cooperation. 
Note that there is also a minimum level of efficiency required to promote native cooperation. This can be seen in the black region of Fig.~\ref{fig.diagrama}$a$, where no matter how large $\mu$ is, native cooperation never emerges. 

The density of native cooperators at fixed $\mu$ does not increase more if $\alpha$ is above a certain value. This is so because the role of the efficient immigrants is local, that is, the benefits of the high efficiency are enjoyed only by the individuals around it. When efficient immigrants arrive, they act as nucleation centers for cooperation clusters by inducing defectors to cooperate. However, for defectors away from nucleation centers, the payoff advantage radiated by clusters of cooperators comes mainly from regular cooperators. 

%

\begin{figure*}

\hspace{0cm}$a)$\hspace{4cm} $b)$ \hspace{4cm} $c)$ \hspace{3.5cm} $d)$

\includegraphics[width=4.2cm]{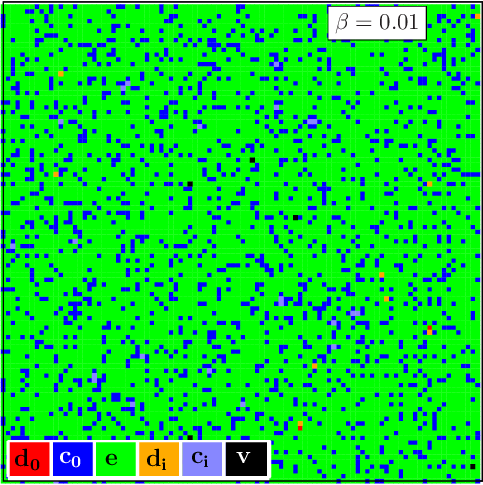}
\includegraphics[width=4.27cm]{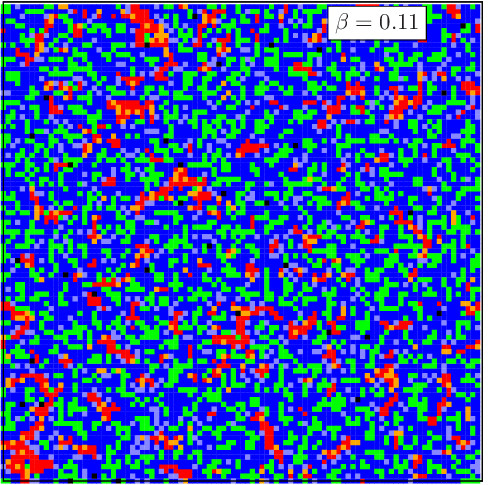}
\includegraphics[width=4.2cm]{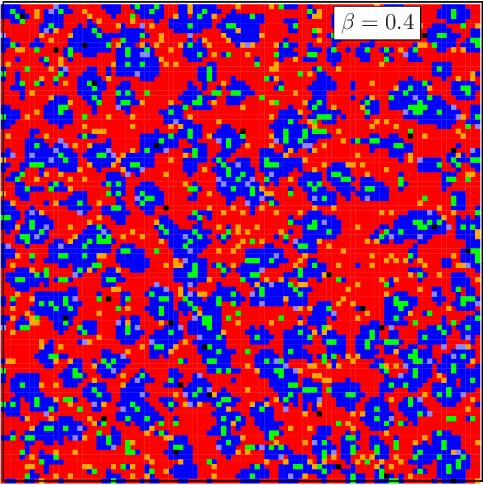}
\includegraphics[width=4.2cm]{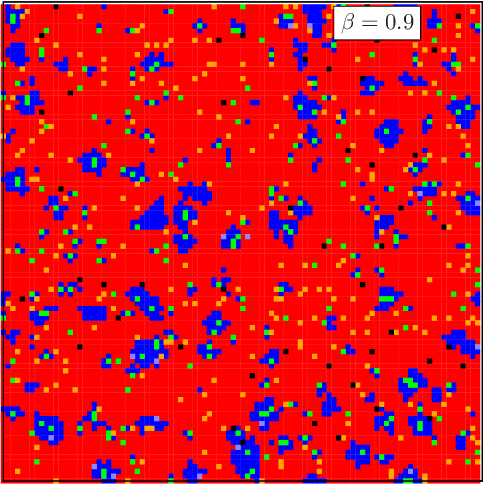}
\caption{Representative spatial configurations at the equilibrium for an efficiency, immigration, and occupation coefficients of $\alpha=4$, $\mu=0.05$, and $\gamma=0.005$, considering different values for the birth coefficient, $\beta$. The figure shows native defectors (red), native cooperators (blue), efficient cooperative immigrants (green), immigrant defectors (orange), immigrant cooperators (light blue), and vacant sites (black). Increasing $\beta$ means increasing the probability of birth and death events in comparison to immigration events. Note that $v$ represents vacant sites.}
\label{fig.snapshot_different_scenarios}
\end{figure*}

\begin{figure*}[htb]

\hspace{0cm}$a)$\hspace{5cm} $b)$ \hspace{5.5cm} $c)$

\includegraphics[width=5.55cm]{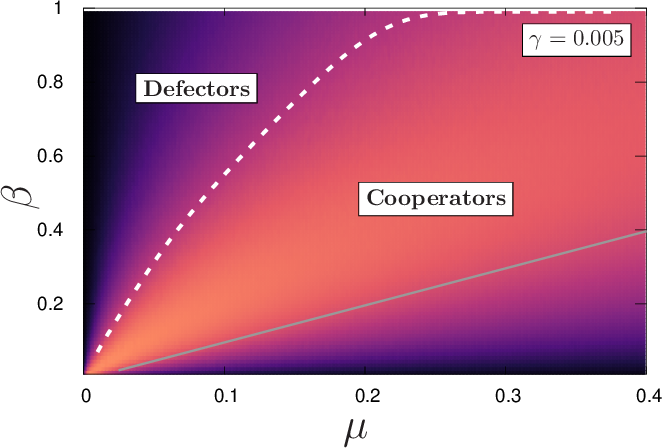}
\includegraphics[width=5.55cm]{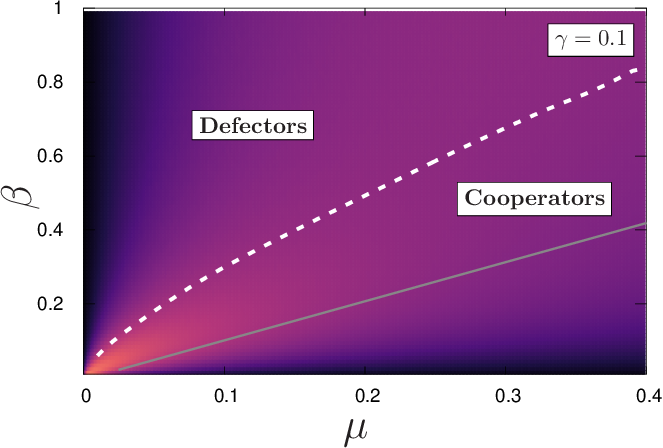}
\includegraphics[width=6.55cm]{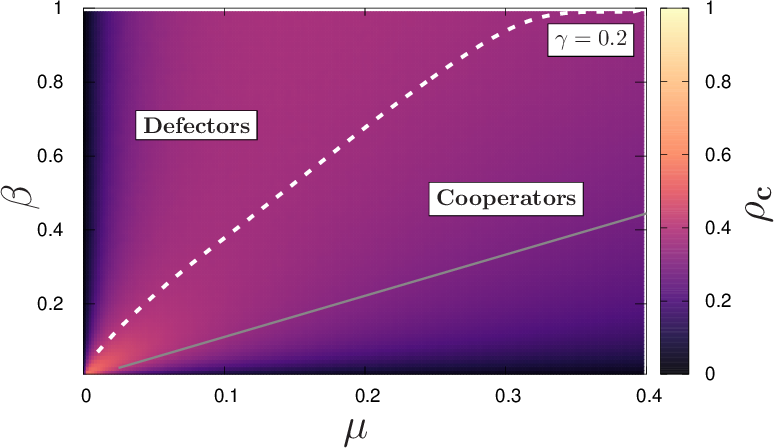}
\caption{Heat map for the equilibrium density of native cooperators as a function of the birth and the immigration coefficients, $\beta$ and $\mu$, considering different values of for the occupation coefficient, $\gamma$. Because systems with different $\gamma$ values have different occupation densities, $\rho_a$, we choose to use a relative density for native cooperators, $\rho_c=C_0/\rho_a$, in order to compare the results. In $a$, $\gamma=0.005$ and the system is almost at its carrying capacity. For $b$ and $c$, $\gamma=0.1$ and $\gamma=0.2$, respectively. For those last values, the number of death events is higher, which generates more available space for immigrants. In all 3 cases the efficiency coefficient considered is $\alpha=4$, while $p_{e}=0.5$. The dashed line separates the region where cooperators, in general, are the majority (below the line), while the grey thin line delimits the area where natives are more frequent (above the line). Notice that the grey line also coincides with the combinations of $\beta$ and $\mu$ where the reproduction probability and the immigration probability are equal, $\beta=\mu/\rho_a$.}
\label{fig.heatmap_betaXmu}
\end{figure*}

On the other hand, for a fixed value of $\alpha$ the native cooperators behave in a non-monotonous way as $\mu$ increases, presenting a peak value of native cooperation for a given optimal $\mu$ (for a broader discussion involving $\beta$ and $\gamma$ see Figs.~\ref{fig.snapshot_different_scenarios} and~\ref{fig.heatmap_betaXmu}). In Fig.~\ref{fig.diagrama}$b$, we show the equilibrium density of native cooperators as a function of the immigration coefficient, $\mu$, for fixed values of efficiency, $\alpha$. First, it is clear that $\alpha$ certainly impacts the outcome. The greater the efficiency of the super-cooperators the higher is the cooperation peak and the lower is the immigration coefficient needed for cooperation to reach such a peak. 
In other words, for a given efficiency level, there is an optimal immigration coefficient that generates a maximum amount of native cooperation. Let $\mu_{cmax}$ be the value of the immigration coefficient at which the peak of the native cooperators density, $C_0^{(max)}$, is reached. The coordinates $(\mu_{cmax},C_0^{(max)})$  are shown in Fig.~\ref{fig.diagrama}$c$ as functions of $1/\alpha$. For visual simplicity, the behaviors of $\mu_{cmax}$ and $C_0^{(max)}$ are consider separately. Interestingly,  $C_0^{(max)}$ presents a linear decrease with $1/\alpha$, tending to its maximum value, $C_0^{(max)}\simeq0.87$, when $\alpha\to\infty$. 
Additionally, the immigration coefficient decreases more rapidly with $\alpha$ than the respective increase in $C_0^{(max)}$, reaching its minimum value, $\mu_{cmax}\simeq0.0156$, when $\alpha\to\infty$. In other words, there must be a minimum influx of immigrants for native cooperation to reach a maximum, even when the cooperative immigrants are infinitely efficient. 
We stress that this tendency is observed for simulations with finite size lattices, but it is observed for all parameters examined.
Also, the existence of an upper bound for $C_0^{(max)}$ that is not equal to unity establishes that, when there is immigration, native cooperators are never able to completely dominate the system. Such limitation is related to native cooperators having to share the space with, at least, the incoming immigrants. Notice, however, that without immigrants, native cooperators would have already disappeared for the set of parameters considered here. 

\begin{figure}[ht]
\includegraphics[width=0.9\linewidth]{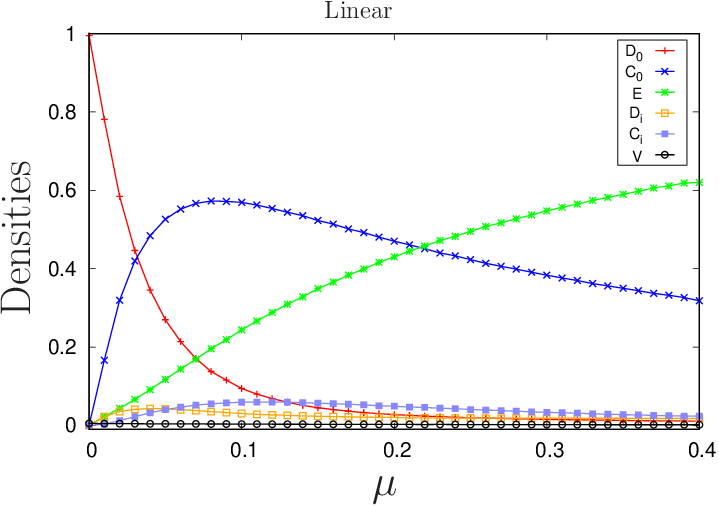}
\caption{Asymptotic densities for the strategies as functions of the immigration coefficient, considering an efficiency of $\alpha=4$, a birth coefficient of $\beta=0.2$, an occupation coefficient of $\gamma=0.005$, and a proportion of efficient cooperators among the incoming immigrants of $p_e=0.5$. Curves $D_0$, $C_0$, $E$, $D_i$, $C_i$ and $V$ correspond to the number of native defectors, native cooperators, efficient immigrant cooperators, immigrant defectors, regular immigrant cooperators, and vacant sites, respectively, normalized by the system size.}
\label{fig.abundances}
\end{figure}

The peak of cooperation observed in Fig.~\ref{fig.diagrama} is due to the initial increase in the native cooperation, which is caused by the efficient immigrants acting as nucleation centers of cooperation, since they are able to outcompete any defector in a sea of defectors when $\alpha>2.78$.
The subsequent decrease in $C_0$ comes from the competition for space. If more immigrants arrive, the competition for space increases, and efficient immigrants become more abundant than native cooperators, as shown in Fig.~\ref{fig.abundances}. The peak is more pronounced the higher the efficiency of the immigrants. Thus, there is an optimal level of immigrants influx if the goal is to promote cooperation among natives without replacing the natives with immigrants.

\begin{figure}[htb]
\begin{center}
\includegraphics[width=0.9\linewidth]{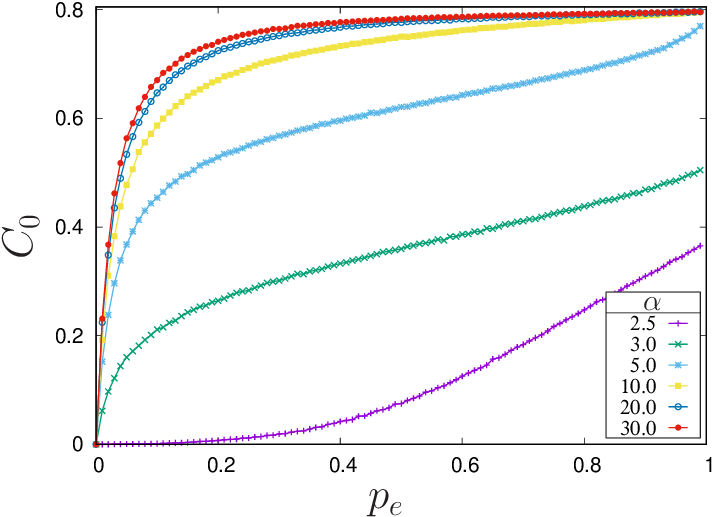}
\end{center}
\caption{Native cooperators population density, $C_0$, as a function of the probability $p_e$ that an incoming immigrant is an efficient cooperator, considering different values of efficiency, $\alpha$. Here we set the immigration and the birth coefficients as $\mu=0.05$ and $\beta=0.2$. It is possible to see that even a tiny fraction of efficient immigrant cooperators greatly increases the final native cooperation fraction.}
\label{fig.densXpce}
\end{figure}

The second remarkable result of the model is that cooperation among natives can be boosted even if the majority of the immigrants are defectors. Figure~\ref{fig.densXpce} shows the average equilibrium density of native cooperators, $C_0$, as a function of the probability that an immigrant is an efficient cooperator, $p_e$.
As expected, $C_0$ is a monotonously increasing function of  $p_e$. Notice, however, that for $\alpha\ge3$ even a small increase in  $p_e$, e.g. $10\%$, is enough to allow native cooperation to flourish.

Interestingly, the effects for different efficiency levels tend to be equivalent near the extremes,  $p_e=0$ and $p_e=1$, especially when $\alpha>3$. This is due to two different processes. For $p_e$ close to $1$, there is a maximum amount of native cooperators allowed in the system for a specific immigration coefficient, as discussed above, which in this case is around $0.8$ (see Fig.~\ref{fig.diagrama}$c$). Thus, more efficient immigrants will have little effect on the final level of cooperation. For $p_e$ close to $0$, on the other hand, the arrival of efficient cooperative immigrants is very rare. 

Additionally, when efficient cooperators arrive in a sea of defectors, their payoff is always greater than their neighbors', as long as $\alpha>2.78$. 
Because of that, efficient cooperators, $e$, act as nucleation centers for the formation of native cooperation clusters. As $e$ is able to provide a better quality contribution, the initial clusters grow sustainably.
We stress that this is an emergent phenomenon, where agents self-organize as cooperators around these centers.
Therefore, the rapid increase in the native cooperators population for small $p_e$ (Fig.~\ref{fig.densXpce}) comes from the formation of $c_0$ clusters around the new nucleation centers arriving more frequently as $p_e$ increases. 
Figures~\ref{fig.snapshot_different_scenarios}$c$ and~\ref{fig.snapshot_different_scenarios}$d$, discussed below, show a couple of examples that illustrate such formation process. 
This rapid increase happens until the $p_e$ fraction is enough to allow all cooperation clusters to percolate. After that, the further increase is due to the few small defector clusters that will be outcompeted. Also, native cooperators are being generated through reproduction, which increases with the number of $e$ in the system.

The positive influence of immigrants depends on the demographics of the population. Figure~\ref{fig.snapshot_different_scenarios} shows the typical effects of increasing the birth and death coefficients using representative lattice snapshots. 
If birth is rare compared to immigration ($\beta<\mu/\rho_a$), the empty sites are occupied mainly by immigrants. The efficient cooperators will initially act as nucleation centers for the formation of native cooperation clusters. However, because reproduction is less frequent than immigration, the population ends up with more immigrants ($e$) than native cooperators. A typical scenario for this configuration is shown in Fig.~\ref{fig.snapshot_different_scenarios}$a$.
Notice, however, that although $e$ may dominate a highly populated system when birth and death are too rare, even in such a scenario native cooperators are able to survive, contrary to what happens when there is no immigration. 

If birth and death events are more frequent than immigration, immigrants are less able to dominate the system, not only because they have to share the empty sites with the offspring of natives, but also because they give birth to natives more frequently, having also to share the empty sites with their offspring, as shown in Fig.~\ref{fig.snapshot_different_scenarios}$b-d$. 
Comparing both situations, we see that some increase in $\beta$, when it is way smaller than $\mu/\rho_a$, may turn the beneficial effect of $e$ more effective, as can be seen in Fig.~\ref{fig.snapshot_different_scenarios}$b$, where $\beta=0.11$. However, if reproduction is way more frequent than immigration, and empty sites are too rare, $d_0$ tends to dominate the system. Those results indicate that there are optimum combinations of the immigration and the reproduction coefficients for the success of native cooperators.
Notice, however, that even when $d_0$ is more frequent, as is the case in Figs.~\ref{fig.snapshot_different_scenarios}$c$ and~\ref{fig.snapshot_different_scenarios}$d$,  $c_0$ may still persist by forming clusters that include a few $e$ as nucleation centers. 

\begin{figure}[htb]
\begin{center}
\includegraphics[width=0.9\linewidth]{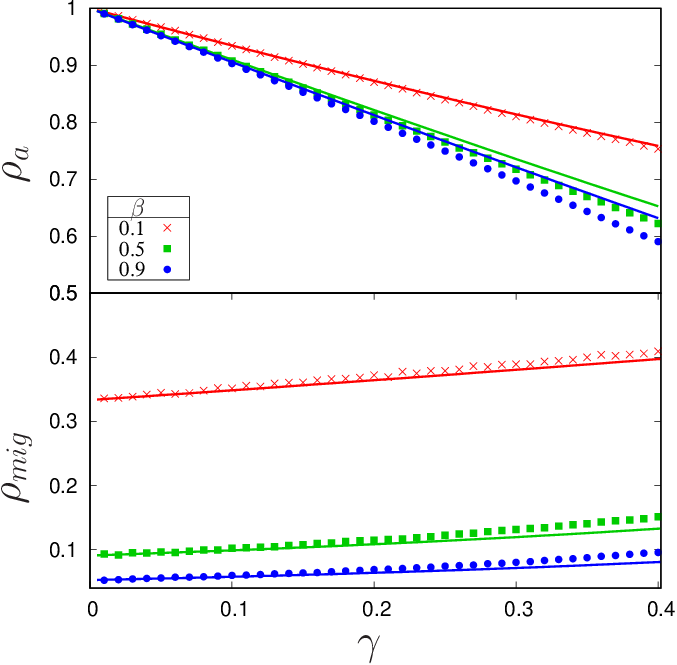}
\end{center}
\caption{Asymptotic occupation density (upper panel) and immigrant density (lower panel), as functions of the occupation coefficient, $\gamma$, for an immigration coefficient of $\mu=0.05$ and 3 different values for the birth coefficient, $\beta$. The lines correspond to the analytical results, while the points are related to the lattice simulations. The two approaches on both figures coincide very well for small $\gamma$.}
\label{fig.analyticXsimul_m005}
\end{figure}

Lastly, we analyze the effect of increasing the frequency of death events relative to birth events, which is controlled by the parameter $\gamma$. A simple mean field analysis (see Appendix) shows that in the limit $\gamma \rightarrow 0$, the density of the active sites approaches one and, if $\mu\ll\beta$, it is approximately given by $\rho_a \approx 1-\gamma$. Thus, small $\gamma$ values correspond to populations close to their carrying capacity. The actual fraction of immigrants in the equilibrium can also be obtained in the mean field analysis, as shown in Fig. \ref{fig.analyticXsimul_m005}. As expected, immigrants are favored at high $\gamma$ values, as there are more empty sites to settle down, and at low $\beta$ values, as competition with reproduction is weakened.

The effect of the demographic parameters  $\beta$, $\mu$ and $\gamma$ on the native cooperators' density is illustrated in Fig.~\ref{fig.heatmap_betaXmu}, where we show the native cooperator density relative to the occupation density, $\rho_c=C_0/\rho_a$, to make the comparison between populations with different $\gamma$ easier.
The peak of native cooperation is observed in all cases. However, the higher the $\beta$, the less pronounced the peak is. The reason is that at high $\beta$ replacement by reproduction and death becomes more important than immigration, as discussed in the last paragraph. 
Interestingly, the system that is most close to its caring capacity, Fig.~\ref{fig.heatmap_betaXmu}$a$, has the highest peak of native cooperators. These results indicate that if the demographics are more stable, the cooperation clusters are also more stable and can resist defection.
We should stress that none of our results depend on the system's initial conditions, as long as the population starts at least partially occupied.

\section{Discussions and conclusion}
To summarize, we developed a simple model that integrates stochastic birth and death dynamics to a public goods game model in the presence of immigration. 
We show that moderate immigration can promote a peak of cooperation among natives in our model.
Our main result suggests that a population in a context that is not favorable for cooperation may greatly benefit from the arrival of just a small fraction of supper cooperative immigrants. 
This is so because when only a few immigrants arrive they act as nucleation centers for native cooperation, influencing defectors to cooperate. The presence of a few efficient cooperators increases the effective multiplicative factor of the few public goods centered around the efficient cooperators, which is enough to increase the power of the cooperative clusters. This boosts local cooperation, which radiates to the population. However, the native cooperation increases only up to a certain level, beyond which the effect of more immigrants is to decrease the number of native cooperators, who compete with immigrants for space.

The demographic parameters determine how beneficial immigration is for the native population. For most of the cases considered here, natives end up being better off with the arrival of immigrants than without it.
When reproduction is rare compared to immigration, defection is washed out, but at the expense of the replacement of natives by immigrants. However, because there can be many defectors among the immigrants, it is still remarkable that cooperation thrives.  
On the other extreme, when reproduction and death are way more frequent than immigration, defectors become more common. Despite that, the small influx of efficient cooperators is still able to prevent defectors from dominating the entire system. 

In the context of evolutionary game theory, the term \textit{migration} usually refers to the ability of an agent to move on the spatial structure where the population is distributed. This approach usually focus on how the mobility, viscosity and the density of players affects cooperation~\cite{Bauza2020, Li2016, fu_jsp13, Funk2019, jiang_ll_pre10, chen_xj_pre12b, ichinose_em08}. 
However, only a few works analyze the impact of the arrival of new individuals,  born somewhere else, to the native community~\cite{jeong2019prisoner, wu2020better}.
These works focus on the issues related to the population capacity to receive immigrants and the immigrants' adaptability to the new society but do not take into account the possible benefits the immigrants' diversity and skills may bring to the native community. 
Here we study how super-cooperative immigrants could benefit the native society through the diversity they introduce into the system. 

We stress that this is a simplified model if compared to real-world immigration, a very complex phenomenon, depending on culture, economics, politics, and many other aspects. Nevertheless, here we aim at an analysis of the emergence of cooperation in a competitive scenario including just the minimal amount of mechanisms to model immigration in a simplified way. This approach highlights the most important aspects of an emergent phenomenon. The proposed model, although based on human migration, can also be further expanded to understand animal migration and individual exchanges between groups with different cooperative traits.

Empirical data shows that immigrants do not free-ride more than natives do~\cite{osili2009immigrants}. However, there is no systematic empirical investigation of how good role models may affect voluntary contributions to the public goods game. Here we see that a few good role models can be enough to boost native cooperation. This work has the potential to serve as a starting point for further empirical investigations, that analyze the effects of immigrants coming from highly cooperative societies.

Our model sheds light on some important global challenges society is facing. When we most need global cooperation, a world of hardening borders is gaining strength, fuelled even more by the 2020 coronavirus pandemic. Interestingly, although creating strong restrictions against immigration may seem like a reasonable action to avoid harm to the native population, our results suggest that, even if defectors come in, as long as a small fraction of efficient cooperators come together, the few supper cooperators may have a way more powerful influence, acting as nucleation centers for the formation of native cooperators clusters.

\begin{acknowledgments}
The authors thank CNPq and Fapemig.
\end{acknowledgments}

\appendix*
\section{Analytical approximation of the occupation and the immigrant/native density}

Considering a mean field approach, some of the demographic aspects of the model can be analyzed through a simple EDO for the temporal variation of the occupation density:
\begin{eqnarray}
\dot{\rho_a}&=&\beta(1-\rho_a)+\mu\left(\frac{1-\rho_a}{\rho_a}\right)-\beta\gamma\nonumber,\\
&=&(\beta\rho_a+\mu)\left(\frac{1-\rho_a}{\rho_a}\right)-\beta\gamma,\label{eq.pa}
\end{eqnarray}
where the two first terms come from reproduction and immigration events, respectively, while the last one is related to death events. Notice that this equation is only valid for a nonempty system, which holds when death is less frequent than the sum of reproduction and immigration events (assuming the system starts populated). The equilibrium points for such equation are those values of $\rho_a$ for which $\dot{\rho}_a=0$~\cite{Brauer2012}. Equation~(\ref{eq.pa}) has one positive and one negative equilibrium points, but because $\rho_a$ is a density, the negative one can never be reached by the system. The positive equilibrium point,
\begin{equation}
    \rho_a^*=\frac12\left[(1-\gamma)-\frac\mu\beta+\sqrt{\left(\frac\mu\beta-(1-\gamma)\right)^2+4\frac\mu\beta}\right],\label{eq.sol_pa}
\end{equation}
depends only on the reproduction, immigration and occupation coefficients, and not on the initial value of $\rho_a$. The sign of $\partial\dot{\rho}_a/\partial\rho_a|_{\rho_a=\rho^*_a}$ determines the stability of $\rho_a^*$~\cite{Brauer2012}. If  $\partial\dot{\rho}_a/\partial\rho_a|_{\rho_a=\rho^*_a}<0$, $\rho_a^*$ is said to be asymptotically stable, and the system asymptotically converges to it (when $\rho_a(t=0)\neq\rho_a^*$). For Eq.~(\ref{eq.sol_pa}), we have:
\begin{equation}
\left.\frac{\partial \dot{\rho}_a}{\partial\rho_a}\right|_{\rho_a=\rho^*_a}=-\left[\beta+\frac{\mu}{(\rho_a^*)^2}\right]<0.
\end{equation}
Thus, $\rho_a^*$ is the asymptotic occupation density for the mean field approach of our model. A comparison between the analytic and the lattice simulation results is shown in Fig.~\ref{fig.analyticXsimul_m005} (top panel) and reveals that both cases coincide well as long as $\gamma$ is small.

Using the occupation density one can also calculate the immigrants' and the natives' density. For the immigrants, the temporal variation of their fraction is given by the following EDO:
\begin{equation}
    \dot{\rho}_{mig}=\mu\left(\frac{1-\rho_a}{\rho_a}\right)-\beta\gamma~\rho_{mig}\label{eq.immig},
\end{equation}
where $\rho_{mig}$ is the immigrants density ($E+D_i+C_i$). The first term in Eq.~(\ref{eq.immig}) comes from the immigration event, while the second one is related to the death of immigrants. This equation has only one nontrivial equilibrium point,
\begin{equation}
    \rho_{mig}^*=\frac{\mu}{\beta\gamma}\left(\frac{1-\rho_a^*}{\rho_a^*}\right),
\end{equation}
which is an asymptotic stable state, since
\begin{equation}
    \left.\frac{\partial \dot{\rho}_{mig}}{\partial\rho_{mig}}\right|_{\rho_{mig}=\rho^*_{mig}}=-\beta\gamma<0.
\end{equation}
The asymptotic natives' density in this mean field approach is given by $1-\rho_{mig}$. Interestingly, such a simple approach well describes the demographic aspects of the lattice version as well, especially for small $\gamma$, as can be seen by the comparison between the mean field and the lattice results in Fig.~\ref{fig.analyticXsimul_m005} (bottom panel).

\end{document}